\documentclass[12pt, authoryear]{elsarticle}

\makeatletter
\def\ps@pprintTitle{%
 \let\@oddhead\@empty
 \let\@evenhead\@empty
 \def\@oddfoot{\centerline{\thepage}}%
 \let\@evenfoot\@oddfoot}
\makeatother

\usepackage{amsmath}
\usepackage{graphicx}
\usepackage{verbatim}
\usepackage{subfigure}
\usepackage{color}
\usepackage{setspace}
\usepackage{float}
\usepackage{natbib}
\usepackage{footnote}
\usepackage{hanging}
\usepackage{url}
\usepackage{lineno}
\makesavenoteenv{tabular}
\makesavenoteenv{table}

\usepackage[top=2.3cm,left=2.3cm,right=2.3cm,bottom=2.3cm]{geometry}

\begin{document}

\begin{frontmatter}

\title{The Eris/Dysnomia system I: The orbit of Dysnomia}

\author[stsci]{B. J. Holler}
\author[low]{W. M. Grundy}
\author[swri]{M. W. Buie}
\author[godd]{K. S. Noll}

\address[stsci]{Space Telescope Science Institute, Baltimore, MD, USA}
\address[low]{Lowell Observatory, Flagstaff, AZ, USA}
\address[swri]{Southwest Research Institute, Boulder, CO, USA}
\address[godd]{Goddard Space Flight Center, Greenbelt, MD, USA}

\begin{abstract}
\doublespacing{We present new results on the Eris/Dysnomia system
  including analysis of new images from the WFC3 instrument on the
  Hubble Space Telescope (HST). Seven HST orbits were awarded to
  program 15171 in January and February 2018, with the intervals
  between observations selected to sample Dysnomia over a full orbital
  period. Using relative astrometry of Eris and Dysnomia, we computed
  a best-fit Keplerian orbit for Dysnomia. Based on the Keplerian fit,
  we find an orbital period of 15.785899$\pm$0.000050 days, which is
  in good agreement with recent work. We report a non-zero
  eccentricity of 0.0062 at the 6.2-$\sigma$ level, despite an estimated
  eccentricity damping timescale of $\leq$17 Myr. Considering the
  volumes of both Eris and Dysnomia, the new system density was
  calculated to be 2.43$\pm$0.05 g cm$^{-3}$, a decrease of $\sim$4\%
  from the previous value of 2.52$\pm$0.05 g cm$^{-3}$. The new
  astrometric measurements were high enough precision to break the
  degeneracy of the orbit pole orientation, and indicate that Dysnomia
  orbits in a prograde manner. The obliquity of Dysnomia's orbit pole
  with respect to the plane of Eris' heliocentric orbit was calculated
  to be 78.29$\pm$0.65$^{\circ}$ and is in agreement with previous
  work; the next mutual events season will occur in 2239. The
  Keplerian orbit fit to all the data considered in this investigation
  can be excluded at the 6.3-$\sigma$ level, but identifying the
  cause of the deviation was outside the scope of this work.}
\end{abstract}

\begin{keyword}
Kuiper belt; Trans-neptunian objects; Hubble Space Telescope observations; Orbit determination
\end{keyword}

\end{frontmatter}

\section{Introduction}
\doublespacing{The Kuiper Belt is a large collection of icy
  bodies found beyond Neptune ($a$$>$30.1 au) that are typically categorized
  into different dynamical populations based on their
  orbital characteristics (e.g., Gladman et
  al., 2008). Many of these Kuiper Belt Objects (KBOs) are thought to
  have formed in other regions of the outer solar system and were
  later emplaced on their current orbits during the era of planetary
  migration (e.g., Malhotra, 1993, 1995; Levison and Morbidelli, 2003;
  Gomes, 2003; Levison et al., 2008). Because of this, a majority of these
  populations contain an assortment of KBOs that span a wide range of
  sizes, colors, and compositions (e.g., M{\"u}ller et al., 2010; Barucci et al., 2011; Brown,
  2012; Hainaut et al., 2012; Lacerda et al., 2014; Bannister et al., submitted). Photometry
  and spectroscopy are the primary tools used to study the
colors and surface compositions of KBOs, with thermal measurements
providing diameter estimates, but the mass of a system can only be
accurately calculated in binary or multiple systems. Binaries are
thought to be quite common among certain
Kuiper Belt populations (Noll et al., 2008, 2014; Fraser et al., 2017), but
they are potentially challenging to detect and characterize due to a combination
of small sizes, low albedos, large heliocentric distances, and small
separations between components.\\
\indent\indent In the case of the dwarf planet (136199) Eris, the
challenge in characterizing the orbit of its satellite Dysnomia is
the system's extreme heliocentric distance ($\sim$96 au), resulting in
a maximum angular separation of $\sim$500 mas as seen from
Earth. Only ground-based facilities equipped with adaptive optics (AO)
and the Hubble Space Telescope (HST) are currently capable of reliably
splitting the two components. The first published orbit for Dysnomia
made use of AO data from Keck and HST data (Brown and Schaller, 2007),
with a more recent orbit fit making use of the same data set plus
previously unpublished HST data obtained in 2015 (Brown and Butler,
2018).\\
\indent Brown and Schaller (2007) report two degenerate orbit solutions
with different values for the semi-major axis of Dysnomia's orbit
(37430$\pm$140 and 37370$\pm$150 km) and period (15.772$\pm$0.002 and
15.774$\pm$0.002 days) that cannot be distinguished from each other
given their uncertainties. These two orbit solutions result in the same
orbit pole obliquity ($\sim$78$^{\circ}$) and two different but equally valid
pole orientations, corresponding to two dates for an orbit opening
angle of 0$^{\circ}$, 2239 and 2126, respectively. Brown and Butler
(2018) report a semi-major axis of 37460$\pm$80 km, which is in
agreement with both solutions from Brown and Schaller (2007). However,
they report a period of 15.78586$\pm$0.00008 days, which differs
significantly from the previously reported periods (6.9-$\sigma$ and
5.9-$\sigma$, respectively). The cause of this discrepancy, given that
the two papers make use of the same data, with Brown and Butler (2018)
only considering two additional data points, is not immediately
clear. Brown and Butler (2018) do not report an orbit pole
obliquity for comparison. Brown and Schaller (2007) initially reported eccentricities of
$<$0.010 and $<$0.013 for the two degenerate orbit solutions,
respectively. Brown and Butler (2018) further constrained the
eccentricity to $<$0.004, suggesting that Dysnomia's orbit is possibly
circular. The combination of previous results on the system mass
(Brown and Schaller, 2007) and radii for Eris (Sicardy et al., 2011)
and Dysnomia (Brown and Butler, 2018) suggest that the system is the
most massive in the Kuiper Belt at (1.66$\pm$0.02)$\times$10$^{22}$ kg, and
has a high estimated density of $>$2.5 g cm$^{-3}$.\\
\indent In this work, we examined relative astrometry of Eris and Dysnomia
in new HST imagery and report updated physical parameters for the system
and updated orbital parameters for Dysnomia. We also report
a pole orientation for Dysnomia's orbit and use this to evaluate the
time of the next mutual events season, when Eris and Dysnomia will
take turns eclipsing each other.}

\section{Observations}
\doublespacing{The imaging observations of Eris and Dysnomia used in
  this work were made between Dec. 3, 2005, and February 3, 2018, with
  NIRC2 at Keck and ACS/HRC, WFPC2/PC1, and WFC3/UVIS on the Hubble Space
  Telescope (HST). We summarize these observations below:
\begin{itemize}
\item Observations with NIRC2 on Keck were carried out as part of three
  different programs in August 2006. The relevant program IDs are C168OL (PI:
  M. Brown, 2006/08/20 \& 2006/08/21), ENG (PI: nirc2eng, 2006/08/30),
  and K240OL (PI: Armandroff, 2006/08/30), and all data are available
  on the Keck Archive. Hundreds of 60-second exposures were taken with
  both the narrow and wide camera settings (plate scales of $\sim$10 and $\sim$40
  mas/pixel, respectively). All observations were made with the laser
  guide star adaptive optics (LGS AO) system in order to separate Eris
  and Dysnomia. We used the published relative astrometry values from
  Brown and Schaller (2007) in this investigation; readers are referred to that paper and
  the associated supplementary material for more information on these data.

\item HST GO programs 10545 and 10860 (PI: M. Brown) observed Eris and
  Dysnomia with the now-defunct High Resolution Channel (HRC) of the
  ACS instrument and the F606W filter. The plate scale of the HRC was
  $\sim$27 mas/pixel and the PSF FWHM was $\sim$50 mas at 0.60
  $\mu$m (as reported in the ACS Instrument
  Handbook\renewcommand{\thefootnote}{\arabic{footnote}}\footnote{\url{https://hst-docs.stsci.edu/display/ACSIHB/}}). Program
  10545 consisted of one HST orbit, referred to as a ``visit,'' of Eris and Dysnomia
  with two nearly consecutive 300-second exposures on
  2005/12/03. Program 10860 consisted of two visits, one with four
  550-second exposures and the other with four 565-second exposures,
  on 2006/08/03. As with the Keck/NIRC2 data described above, the
  relative astrometry values used in this investigation were taken
  directly from Brown and Schaller (2007).

\item HST GO program 11169 (PI: M. Brown) observed Eris and Dysnomia
  with the now decommissioned Wide Field and Planetary Camera 2
  (WFPC2) on 2007/08/13. All 4 images were made in Visit 13 using
  Planetary Camera 1 (PC1). Two of the images were taken through the F606W
  filter with an exposure duration of 400 seconds; the other two images were
  500 seconds and were taken through the F814W filter. The plate scale of
  WFPC2/PC1 was 46 mas/pixel. These data have not been previously
  published.

\item HST GO program 13668 (PI: M. Buie) consisted of 2 nearly
  consecutive visits on both 2015/01/29 and 2015/02/01, for 4 total
  visits. Each visit consisted of one 80-second exposure and three
  720-second exposures with WFC3/UVIS and the F350LP filter. The plate
  scale of WFC3/UVIS is 40 mas/pixel, with a PSF FWHM of $\sim$72 mas
  at 0.35 $\mu$m (as reported in the WFC3 Instrument
  Handbook\renewcommand{\thefootnote}{\arabic{footnote}}\footnote{\url{https://hst-docs.stsci.edu/display/WFC3IHB}}). Astrometry
  of Eris and Dysnomia from these observations was previously published in Brown
  and Butler (2018); other observations from this program were used to
  identify the satellite of Makemake (Parker et al., 2016).

\item HST GO program 15171 (PI: B. Holler) consisted of 7 visits
  made between 2018/01/01 and 2018/02/03. Each orbit consisted of four
  348-second exposures and one 585-second exposure with WFC3/UVIS and
  the F606W filter. The PSF FWHM is $\sim$67 mas at 0.60 $\mu$m (as
  reported in the WFC3 Instrument Handbook). These
  visits were originally planned to occur within one orbital period of
  Dysnomia ($\sim$16 days; Brown and Schaller, 2007). However, visit 3
  (2018/01/03) was subject to a tracking failure, and so two of the
  348-second exposures and the 585-second exposure were streaked and
  not used in this analysis. An additional visit was awarded on
  2018/02/03 to compensate for these losses. Median images from all
  visits that were not subject to tracking errors are presented in
  Figure 1 and show the relative positions of Eris and Dysnomia. These
  are new observations that have not been published previously.
\end{itemize}

The NIRC2, ACS/HRC, and WFPC2/PC1 data sets were considered together
as one epoch (2005-2007) and the WFC3/UVIS data sets were considered as a separate
epoch (2015-2018) for the following analysis. We refer to these as Epoch 1 and
Epoch 2, respectively.}

\begin{figure}[h!]
\begin{center}
\includegraphics[scale=0.72]{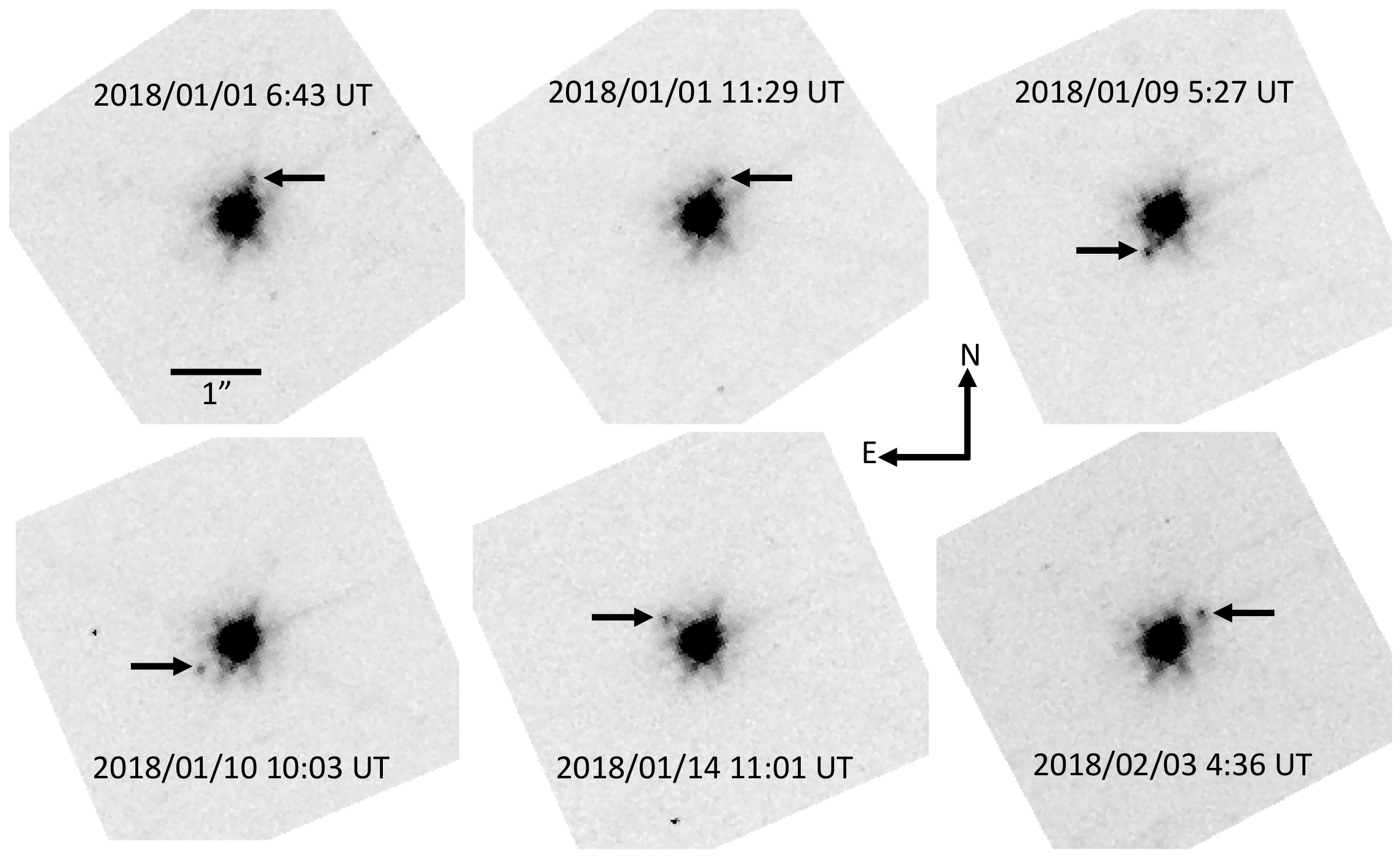}
\caption{Median of the four 348-second images from six visits of HST
  15171 stretched to show both Eris and Dysnomia (denoted by the
  arrow). All images are shown using the same
  stretch and are rotated so that North is up and East is
  to the left. The median UT date and time
are given for each image. Visits 1, 2, \& 4 are along the top row; visits 5, 6, \& 53
are along the bottom row. Visit 3 consisted of only two usable images
so we do not present the median image here.}
\end{center}
\end{figure}

\section{Analysis \& Results}
\doublespacing{Relative astrometry of Eris and Dysnomia for Epoch 1
  was taken primarily from the supplementary information of Brown and Schaller
  (2007). The WFPC2 data from HST program 11169, included in Epoch 1 and
  presented here for the first time, were reduced with the
 WFPC2 pipeline, \textit{calwp2} v2.5.5 (released April 17, 2009)\renewcommand{\thefootnote}{\arabic{footnote}}\footnote{\url{http://documents.stsci.edu/hst/wfpc2/documents/handbooks/dhb/wfpc2_dhb.pdf}}. The
 \textit{calwp2} pipeline performs an analog-to-digital correction,
 marks bad pixels, subtracts the bias image and dark frame,
 performs flat fielding, and applies a shutter shading
 correction. The relative positions of Eris and Dysnomia were then
 extracted from these processed images.\\
\indent Relative astrometry of Eris and Dysnomia for Epoch 2 was
performed on the reduced \textit{*\_flt.*} images retrieved from the MAST archive at
  STScI. Reduction of the raw images from HST programs 13668 and
  15171 was handled by the WFC3 pipeline, \textit{calwf3} v.3.4.1 (released April 10,
2017)\renewcommand{\thefootnote}{\arabic{footnote}}\footnote{\scriptsize{\url{http://www.stsci.edu/files/live/sites/www/files/home/hst/instrumentation/wfc3/_documents/wfc3_dhb.pdf}}}. The
\textit{calwf3} pipeline flags bad pixels, performs bias
correction, trims overscan regions, subtracts the contribution from
the dark current, performs flat fielding, and normalizes the fluxes between the
UVIS1 and UVIS2 detectors. The \textit{*\_flt.*} files do not undergo
charge transfer efficiency (CTE) correction and no post-flash or
cosmic ray corrections were implemented.\\
\indent We used the method of PSF-fitting to calculate the positions of
  the Eris and Dysnomia PSFs to sub-pixel precision for the WFPC2 and
  WFC3 data. Tiny Tim version 7.5, the most recent version of the HST PSF-simulation
  software (e.g., Krist, 1993), was used to construct a grid of PSFs
  across the WFPC2 and WFC3 images. The PSF varies across the image
  due to distortion and the grid was sufficiently sampled to
  account for this effect. An initial guess for the position of Eris
  in the image was made to determine the nearest PSF on the grid for
  use in the PSF-fitting. KBOs are redder than the Sun, so we assumed
  the color index of a G5V star when creating the PSFs in Tiny Tim. Eris and
  Dysnomia were fit simultaneously using the \textit{amoeba} IDL
  routine. The output centroid positions in detector coordinates were
  converted to right ascension and declination using the \textit{xyad}
  IDL routine and the WCS solution provided in the FITS
  headers. Offsets of Dysnomia from Eris were calculated by
  subtracting the locations of the two objects. The right ascension
  and declination offsets and their corresponding 1-$\sigma$ uncertainties
  are presented in Table 1. Uncertainties are reported as the standard
  deviation of the offsets from the individual images of each
  visit. An ``error floor'' was set at 2.26 mas, which was the average
  astrometric uncertainty in both right ascension and declination
  across all of the individual images from Epoch 2.

\begin{table}[h!]
\begin{center}
\textbf{Table 1}\\
Offsets of Dysnomia from Eris\\
\small
\begin{tabular}{lccc}
\hline
Mean UT date & RA offset (mas) & Dec offset (mas) & \# of images\\
\hline
\textit{Epoch 1} & & & \\
2007/08/13 2.04194 & 484.67$\pm$4.15 & -178.10$\pm$2.68 & 4\\
& & & \\
\textit{Epoch 2} & & & \\
2015/01/29 4.02507 & -350.03$\pm$2.26 & -226.03$\pm$2.26 & 4\\
2015/01/29 5.70083 & -341.39$\pm$2.26 & -229.13$\pm$2.26 & 4\\
2015/02/01 10.02840 & +287.34$\pm$2.26 & -328.08$\pm$2.26 & 4\\
2015/02/01 11.72611 & +297.64$\pm$2.26 & -323.68$\pm$2.26 & 4\\
2018/01/01 6.72108 & -128.18$\pm$3.67 & +363.10$\pm$2.26 & 5\\
2018/01/01 11.48714 & -171.78$\pm$4.74 & +362.48$\pm$2.26 & 5\\
2018/01/03 6.18291 & -454.62$\pm$5.95 & +226.35$\pm$3.61 & 2\\
2018/01/09 5.45492 & +145.52$\pm$3.59 & -366.50$\pm$2.74 & 5\\
2018/01/10 10.06408 & +374.03$\pm$2.26 & -296.19$\pm$2.32 & 5\\
2018/01/14 11.00769 & +365.82$\pm$2.26 & +221.95$\pm$2.26 & 5\\
2018/02/03 4.60881 & -378.30$\pm$2.26 & +277.49$\pm$2.26 & 5\\
\hline
\end{tabular}
\end{center}
\end{table}

\indent The offsets of Dysnomia from Eris, along with geocentric distance and
mean UT date, were used to compute the best-fit Keplerian orbit of
Dysnomia around Eris shown in Figure 2, with the residuals presented
in Figure 3. We fit the following set of parameters (see Table 2 for
variable definitions): $P$, $a$, $e$cos$\omega$, $i$, $\epsilon$,
$\Omega$, $e$sin$\omega$. The eccentricity, $e$, and argument of
periapsis, $\omega$, were extracted from $e$cos$\omega$ and
$e$sin$\omega$. The eccentricity was calculated by adding $e$cos$\omega$ and
$e$sin$\omega$ in quadrature, then the argument of periapsis,
$\omega$, was calculated using the eccentricity and either $e$cos$\omega$ or
$e$sin$\omega$. The longitude of perihelion, $\varpi$, is defined
as $\omega$+$\Omega$ and is reported in Table 2. The  Keplerian orbits
were computed for Epoch 1 and
Epoch 2, separately, and Epochs 1 \& 2 combined. The best-fit values,
along with the $\chi^2$ for each fit, are reported in Table 2. We
adopt the values from the combined fit for the orbit of Dysnomia.\\
\indent To calculate the uncertainties on the fitted parameters, we
took each set of nominal parameters and varied each separately around
the best-fit value, calculating the $\chi^2$ for the orbit at each step. By varying
only one parameter at a time the other parameters were forced to
adjust. A parabola was then fit to $\chi^2$ as a
function of parameter value. The best-fit value from the orbit fit
corresponded to the minimum $\chi^2$ and asymmetric errors were found where
the parabolic fit was equal to $\chi^2_{min}$+1. The uncertainties reported in
Table 2 for the fitted elements are symmetric and are equal to the
larger of the two asymmetric errors.

\begin{table}[h!]
\begin{center}
\textbf{Table 2}\\
Orbital parameters and 1-$\sigma$ uncertainties for epoch 2453979.0 JD\\
\scriptsize
\begin{tabular}{llccc}
\hline
Parameter & & Epoch 1 & Epoch 2 & Combined \\
& &  ($\chi^2$=18.9) & ($\chi^2$=69.6) & ($\chi^2$=107.6) \\
\hline
\textit{Fitted elements} & & & & \\
Period (days) & \textit{P} & 15.78674$\pm$0.00092 & 15.78645$\pm$0.00019 & 15.785899$\pm$0.000050\\
Semi-major axis (km) & \textit{a} & 37636$\pm$216 & 37332$\pm$94 & 37273$\pm$64\\
Eccentricity & \textit{e} & 0.0156$\pm$0.0059 & 0.0035$\pm$0.0011 & 0.0062$\pm$0.0010\\
Inclination$^a$ (deg) & \textit{i} & 45.87$\pm$0.88 & 45.32$\pm$0.18 & 45.49$\pm$0.15\\
Mean longitude at epoch (deg) & $\epsilon$ & 124.8$\pm$1.0 & 128.75$\pm$0.83 & 125.78$\pm$0.32\\
Longitude of ascending node$^a$ (deg) & $\mathit{\Omega}$ & 126.2$\pm$1.1 & 126.16$\pm$0.32 & 126.17$\pm$0.26\\
Longitude of periapsis$^a$ (deg) & $\varpi$ & 28$\pm$20 & 322$\pm$24 & 307$\pm$12\\
\\
\textit{Derived parameters} & & & & \\
Standard gravitational parameter (km$^3$ s$^{-2}$) & $\mu$ & 1131$\pm$19 & 1104$\pm$8 & 1099$\pm$6\\
System mass (10$^{22}$ kg) & $M_{sys}$ & 1.695$\pm$0.029 & 1.654$\pm$0.012 & 1.6466$\pm$0.0085\\
Orbit pole right ascension$^a$ (deg) & $\alpha_{pole}$ & 36.2$\pm$1.1 &36.16$\pm$0.32  & 36.17$\pm$0.26\\
Orbit pole declination$^a$ (deg) & $\delta_{pole}$ & 44.13$\pm$0.88 & 44.68$\pm$0.18 & 44.51$\pm$0.15\\
Orbit pole ecliptic longitude$^b$ (deg) & $\lambda_{pole}$ & 48.98$\pm$0.91 & 49.18$\pm$0.25 & 49.12$\pm$0.21\\
Orbit pole ecliptic latitude$^b$ (deg) & $\beta_{pole}$ & 28.05$\pm$0.87 & 28.57$\pm$0.19 & 28.41$\pm$0.16\\
Inclination to heliocentric orbit (deg) & $i_{helio}$ & 78.47$\pm$0.79 & 78.21$\pm$0.66 & 78.29$\pm$0.65\\
Next mutual events season (year) & & 2239 & 2239 & 2239\\
\hline
$^a$Referenced to J2000 equatorial frame.\\
$^b$Referenced to J2000 ecliptic frame.
\end{tabular}
\end{center}
\end{table}

\begin{figure}[h!]
\begin{center}
\includegraphics[scale=0.8]{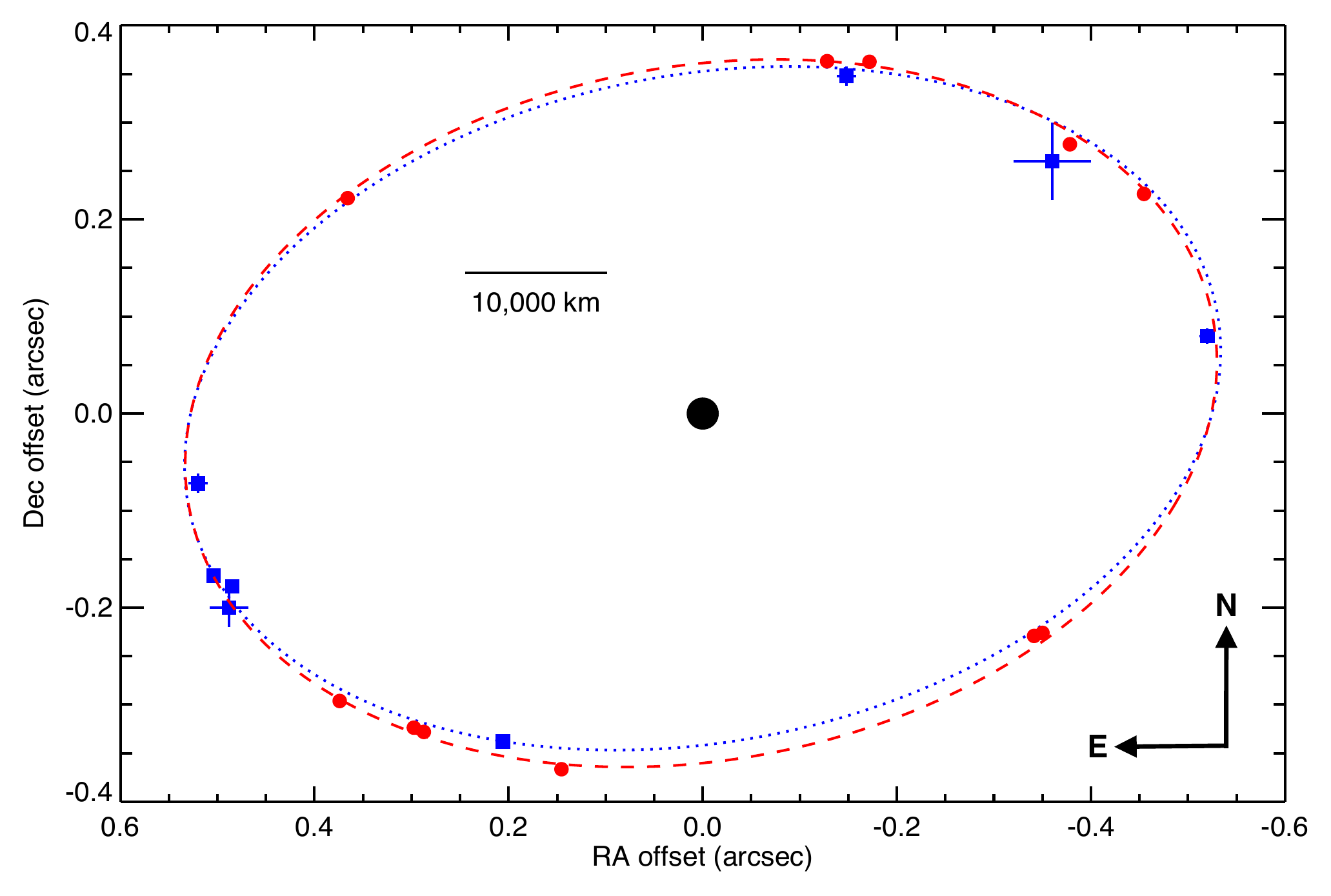}
\caption{Projected orbit of Dysnomia. North is up and East is to the
  left, in the direction of increasing right ascension. Eris is
  to-scale in the center ($\sim$30 mas diameter). The blue squares represent the positions
  of Dysnomia from Epoch 1. The red circles represent the positions of
  Dysnomia in Epoch 2. These symbols are not scaled to the
  estimated diameter of Dysnomia. Error bars are shown for all points
  but may be smaller than the symbol (see the supplementary material
  from Schaller and Brown (2007) for errors for Epoch 1 and Table 1
  for Epoch 2). The blue dotted and red dashed lines
  represent the orbit fits to Epoch 1 and Epoch 2, respectively.}
\end{center}
\end{figure}

\begin{figure}[h!]
\begin{center}
\includegraphics[scale=0.8]{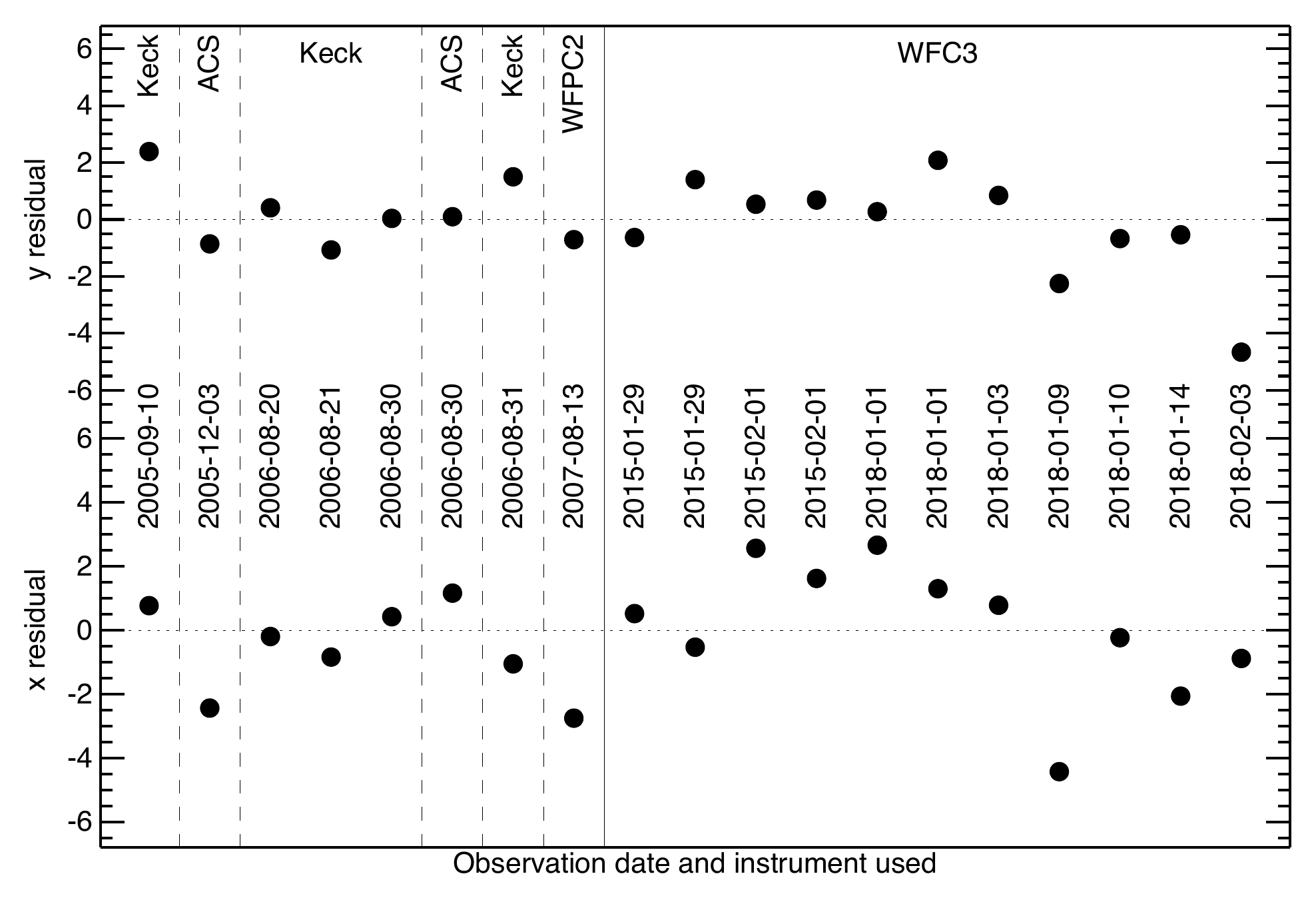}
\caption{Residuals in $x$ (right ascension) and $y$ (declination) for each
  observation. Units are number of 1-$\sigma$ error bars. Each
  observation is labeled with the name of the instrument at the top of
  the plot and the UT date vertically in the middle. The solid grey
  line separates the Epoch 1 residuals (left side) from the Epoch 2
  residuals (right side).}
\end{center}
\end{figure}

\indent The derived parameters (and their uncertainties) in Table 2
were calculated using the fitted elements (and their
uncertainties). The system mass, $M_{sys}$, was calculated from the
period, $P$, and the semi-major axis, $a$, of Dysnomia using Newton's
version of Kepler's Third Law. The standard gravitational parameter is
simply the system mass multiplied by the gravitational constant. The
equatorial coordinates of the Dysnomia orbit pole were
calculated as $\alpha_{pole}$=$\Omega$ -- 90$^{\circ}$ and
$\delta_{pole}$=90$^{\circ}$ -- $i$. Due to the projection
of the orbit onto the sky, there were two possible solutions for the
orbit pole orientation. To break this mirror degeneracy, we
evaluated the orbit solution for both cases and report the values for
the orbit with the lower $\chi^2$. This orbit pole position was then
converted to ecliptic coordinates ($\lambda_{pole}$, $\beta_{pole}$)
using standard spherical trigonometric formulae. The
inclination to heliocentric orbit is the angle between Dysnomia's orbit
pole and heliocentric orbital pole, and is also referred to as the
obliquity of the orbit. It was calculated by taking the dot product
between the Eris ($\lambda_{Eris}$=305.95$^{\circ}$,
$\beta_{Eris}$=45.99$^{\circ}$) and Dysnomia orbit pole vectors. A value
$<$90$^{\circ}$ indicates that Dysnomia's orbit is prograde.\\
\indent The opening angle of Dysnomia's orbit between 1600 and 2500
C.E. was calculated using the orbit pole orientation determined in
this work and a vector table of Cartesian positions from
JPL/Horizons. The vector table for Eris was calculated with respect to
the solar system barycenter in 1-year timesteps and included the light travel time
correction. At each timestep, the $x$, $y$, and $z$ positions of Eris
defined the Eris-Sun vector (the distance from the center of
the Sun to the solar system barycenter is negligible compared to the
distance between Eris and the Sun). The orbit pole vector was defined
in Cartesian coordinates using the orbit pole ecliptic latitude
($\beta_{pole}$) and longitude ($\lambda_{pole}$) from Table 2 and the
equations below:

\begin{align}
x=\mathrm{cos}(\beta_{pole})\mathrm{sin}(\lambda_{pole})\nonumber\\
y=\mathrm{cos}(\beta_{pole})\mathrm{sin}(\lambda_{pole})\nonumber\\
z=\mathrm{sin}(\beta_{pole})\nonumber
\end{align}

\noindent Taking the dot product of the Eris-Sun vector and Dysnomia's orbit pole vector
yielded the value of the opening angle at that timestep. The
orbit opening angle between 1600 and 2500 C.E. is presented in
Figure 4. The opening angle in early 2018 was 42$^{\circ}$, and the
next mutual events season, when the opening angle reaches 0$^{\circ}$,
will occur in 2239.}

\begin{figure}[h!]
\begin{center}
\includegraphics[scale=0.75]{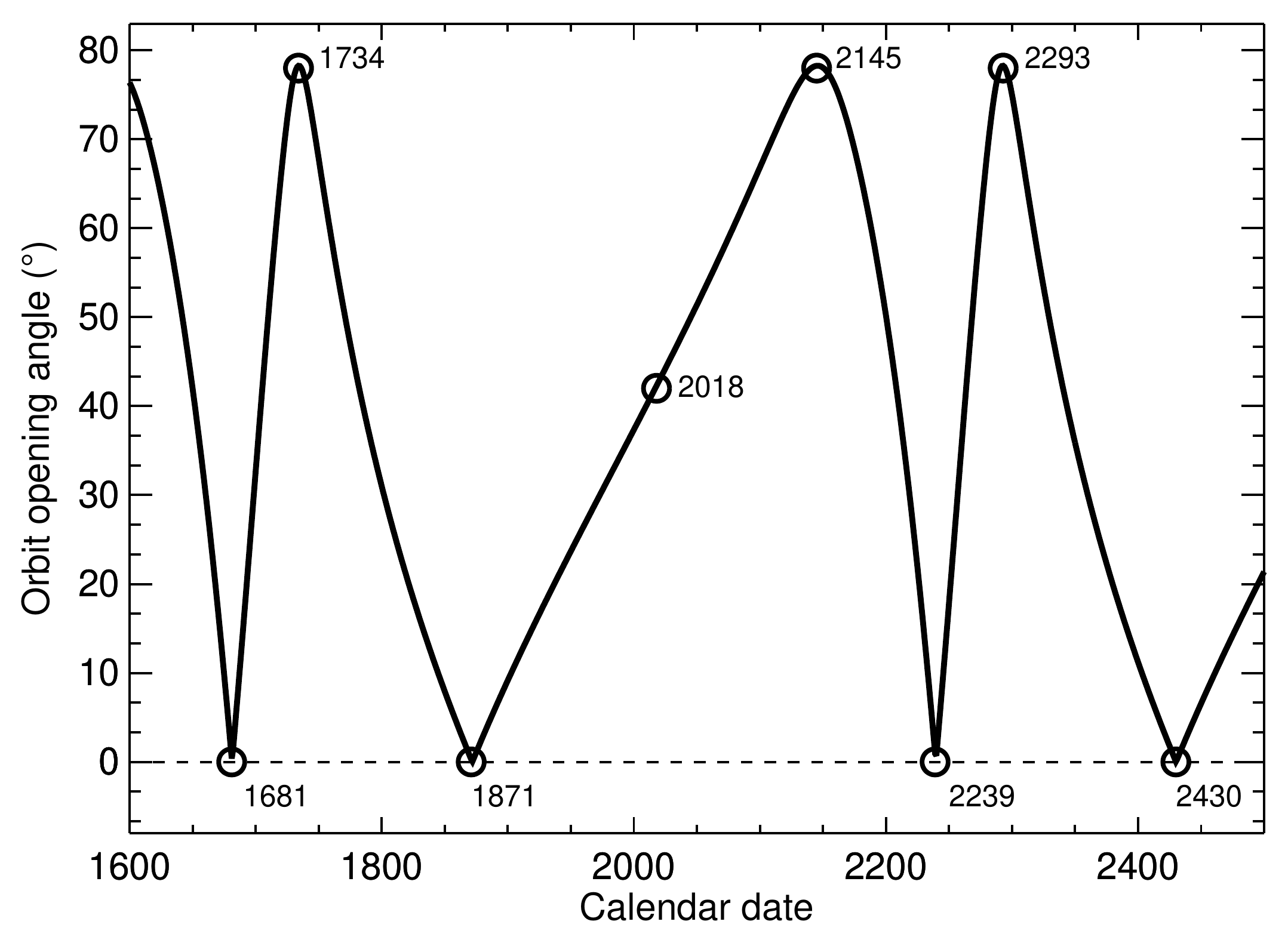}
\caption{Opening angle of Dysnomia's orbit as a function
  of date from 1600 to 2500 C.E; Eris' orbital period is approximately
  558 Earth-years. The dates of maximum and minimum
  opening angles, as well as the most recent HST
  observations, 2018, are marked with open black circles. Mutual
  events occur when the opening angle reaches 0$^{\circ}$, with the
  next instance in 2239. Assuming different values for $\beta_{pole}$ and
  $\lambda_{pole}$ (within the 1-$\sigma$ error bars) changes the dates
  in this figure by no more than a few months and the opening angle by
  no more than $\sim$0.20$^{\circ}$ on any given date.}
\end{center}
\end{figure}

\section{Discussion}
\doublespacing{The orbit pole obliquity (78.29$\pm$0.65$^{\circ}$) and
  the next mutual events season (2239 C.E.) determined in this work are
  consistent with the values calculated for Orbit 1 in Brown and Schaller
  (2007). The semi-major axis and system mass from the combined fit of
  this work are also in agreement with Orbit 1 (within 1-$\sigma$);
  the semi-major axis is in agreement with the value reported in
  Brown and Butler (2018) to within 2-$\sigma$. The period
  calculated in this work is in very good agreement with that reported
  in Brown and Butler (2018), with a difference of only $\sim$3 seconds, but
  both values are inconsistent with those from Brown and Schaller
  (2007): 15.772$\pm$0.002 days for Orbit 1 and 15.774$\pm$0.002 days
  for Orbit 2. This is surprising given that the semi-major axis, period, and
  system mass are all related through Newton's version of
  Kepler's Third Law; however, the Brown and Schaller (2007) results
  for the period are clearly the outliers. Additionally, we note that Epoch
  1 from this work, which made use of the same relative astrometry as
  Brown and Schaller (2007), plus an additional WFPC2 measurement, has
  a period which is in agreement with the fit to the Epoch 2 data, but
  not with the results of Brown and Schaller (2007).\\
\indent The orbit angles (inclination and longitude of the ascending
node) reported in both Brown and Schaller (2007) and Brown and Butler
(2018) were referenced to the J2000 ecliptic frame, whereas those same
values in this work were referenced to the J2000 equatorial frame
(Table 2). After converting from the equatorial to the ecliptic frame,
we found an inclination of 61.59$\pm$0.14$^{\circ}$ and a longitude of the
ascending node of 139.12$\pm$0.22$^{\circ}$. (We adopted the
uncertainties from the equatorial frame for the parameters in the
ecliptic frame.) We find that the inclination
and longitude of the ascending node reported in this work agree with
those for Orbit 1 from Brown and Schaller (2007),
61.3$\pm$0.7$^{\circ}$ and 139$\pm$1$^{\circ}$, to within 1-$\sigma$
and with those from Brown and Butler (2018), 61.1$\pm$0.3$^{\circ}$
and 139.6$\pm$0.2$^{\circ}$, to within 2-$\sigma$.\\
\indent We note a discrepancy in the RA offset between this work
(Table 1) and Brown and Butler (2018) for 2015/02/01. Both
investigations performed relative astrometry on images obtained as
part of HST GO program 13668 (PI: M. Buie), and both investigations
used a PSF-fitting method with PSFs generated by Tiny Tim. In this
work, we report relative astrometry between Eris and Dysnomia for each
visit of program 13668, whereas Brown and Butler (2018) reported a
single RA offset and a single Dec offset for each UT date. Taking the
average of our RA offsets and Dec offsets, in mas, for each
UT date resulted in values of (-345.71$\pm$3.20, -227.58$\pm$3.20) for
2015/01/29 and (292.49$\pm$3.20, -325.88$\pm$3.20) for
2015/02/01. Comparison of these offsets to those in Brown and Butler
(2018) resulted in differences, in mas, of (1.29$\pm$3.77,
1.58$\pm$3.35) for 2015/01/29 and (10.49$\pm$4.39, 0.88$\pm$3.77) for
2015/02/01. The offsets differ by less than 1-$\sigma$ except for the
RA offset on 2015/02/01, which shows a 2.4-$\sigma$ difference. The
difference is $\sim$10.5 mas, which corresponds to about a quarter of
a WFC3 pixel (40 mas/pixel plate scale) and about 15\% of the PSF
FWHM. This difference is therefore at the sub-pixel level and appears not to
have produced a noticeable difference in
the orbit fits between the two investigations. This discrepancy may
simply be due to differences in the exact implementation of the
PSF-fitting methods, such as the PSF color index and/or the
detector location at which the PSF was calculated.\\
\indent The density of Eris was originally calculated to be
2.3$\pm$0.3 g cm$^{-3}$ using the mass from Brown and Schaller (2007)
and a radius of 1200$\pm$50 km determined directly from HST images
(Brown et al., 2006). A stellar occultation by Eris in 2010 resulted
in a more precise radius measurement of 1163$\pm$6 km, and the density
was revised upwards to 2.52$\pm$0.05 g cm$^{-3}$ (Sicardy et al.,
2011). Brown and Butler (2018) reported a radius for Dysnomia of
350$\pm$57.5 km from thermal observations, which makes Dysnomia a large
KBO in its own right and its contribution to the system
mass therefore cannot be ignored. Using our more precise mass
estimate of (1.6466$\pm$0.0085)$\times$10$^{22}$ kg, the Eris radius
from Sicardy et al. (2011), and the Dysnomia
radius from Brown and Butler (2018), we calculate a system density of
2.43$\pm$0.05 g cm$^{-3}$. The only reliable density measurement for a
KBO satellite is for Charon, which has a density roughly 92\% of
Pluto's density (Stern et al., 2015); thus, it is not out of the
question for Dysnomia to have a density which is comparable to
Eris'. If Eris and Dysnomia have the same density, 2.43 g cm$^{-3}$,
then Dysnomia accounts for $\sim$3\% of the total mass of the
system. If Dysnomia has a much lower density than Eris (e.g., 0.8 g
cm$^{-3}$) then it accounts for $<$1\% of the system mass.\\
\indent Orbit 1 from Brown and Schaller (2007) constrained Dysnomia's
eccentricity to $<$0.010 and Brown and Butler (2018) further reduced
this upper limit to $<$0.004. The eccentricity of the combined fit
reported in this work, 0.0062$\pm$0.0010, is nominally in agreement
with the Brown and Butler (2018) result, given the uncertainty
reported in this work. The takeaway from the consistency between these
two values is that Dysnomia's orbital eccentricity is exceedingly
low. However, based on
the uncertainty on the eccentricity measurement, we can
report that the eccentricity is non-zero at the 6.2-$\sigma$
level. We calculated the timescale for tidal circularization based on
the nominal radius of Dysnomia (350 km; Brown and Butler, 2018) and
two extreme values for Dysnomia's density (0.8 and 2.43 g
cm$^{-3}$) using the equation from Goldreich and Soter (1966):
\begin{equation}
\tau_e=\frac{4}{63}Q \frac{M_D}{M_E}
\left(\frac{a^3}{GM_E}\right)^{1/2}\left(\frac{a}{R_D}\right)^5.\nonumber
\end{equation}
In the above equation, $Q$ is the unitless tidal dissipation factor
(typically assumed to be 100
in the absence of additional information), $M_D$ is the mass of Dysnomia,
$M_E$ is the mass of Eris and taken to be $M_{sys}-M_D$, $a$ is
the semi-major axis from the combined fit, and $R_D$ is the radius of
Dysnomia. For densities of Dysnomia of 0.8 and 2.43 g cm$^{-3}$ we
computed circularization timescales of $\sim$5 Myr and $\sim$17 Myr,
respectively. Thus, regardless of Dysnomia's density, it should be on
a perfectly circular orbit given its current semi-major axis.\\
\indent The non-zero eccentricity could be real and a result of
Dysnomia being in resonance with an unseen interior satellite. It is
also possible that the eccentricity is not real and is instead a
result of center-of-light versus center-of-body (CoL-CoB) offsets or
systematic errors. A CoL-CoB effect is a result of large-scale,
potentially high-contrast, albedo patterns and offsets a measured PSF centroid
away from the actual center of the body because lower-albedo regions
account for a smaller fraction of the PSF flux. The difference between
Dysnomia's periapse and apoapse is 462$\pm$105 km, which is comparable
to Dysnomia's radius (Brown and Butler, 2018). Considering the extreme case of an
Iapetus-like Dysnomia with a large hemispherical contrast in visible
albedo, the CoL-CoB offsets could be due to Dysnomia alone. However,
Dysnomia's light curve amplitude is currently unconstrained and
Iapetus' two-tone coloration appears to be a unique case made possible
by Saturn's dust environment (e.g., Spencer and Denk,
2010). Additionally, Eris has a very low light curve amplitude that
could be due to a uniform albedo distribution across
the surface (e.g., Carraro et al., 2006; Duffard et al., 2008; Roe et
al., 2008). If this is the case, Eris would not contribute appreciably
to any CoL-CoB offsets. If instead we are viewing Eris nearly pole-on,
the same regions of the surface would always be observed and would
explain the low light curve amplitude, but would not preclude
large-scale albedo features from creating a CoL-CoB offset. Thus,
CoL-CoB offsets on both Eris and Dysnomia could combine to produce the
non-zero eccentricity, but this is speculation and no firm evidence
exists supporting this interpretation.\\
\indent Another option is that the relative astrometric measurements
are subject to systematic offsets, though it is unclear where they
originated from. For instance, such offsets would not be due to
the motion of Dysnomia during each 348-second exposure. At 96.6 au,
one WFC3/UVIS pixel (40 mas) corresponds to $\sim$2800 km and Dysnomia
moves through 0.026\% of its orbit, or $\sim$30 km, in 348
seconds. Thus, Dysnomia's apparent motion shifts it, at most, $\sim$1\% of a
pixel in each exposure, depending on the position along its projected
orbit, and this is less than the uncertainty on the relative
astrometry. It also seems unlikely that the offsets arise from the
non-radially symmetric PSF of Eris affecting the PSF fit to
Dysnomia. Not only should this be accounted for by fitting the PSFs to
both objects simultaneously, but Dysnomia's PSF is not always affected
by the same structures in Eris' PSF as it moves through different orbital longitudes. A
somewhat more realistic possibility is that the color correction
applied to the PSFs produced by Tiny Tim is not equally applicable to
Eris and Dysnomia. We did not test different color corrections for the
Tiny Tim PSFs so it is unclear if this would have a preferential and
appreciable effect on the central position of either the Eris or
Dysnomia PSF in every image. There is, of course, the possibility that a
systematic effect is at play that we did not identify here. However, without
of a well-supported and more believable alternate solution, we accept
the non-zero eccentricity and 6.2-$\sigma$ significance at
face-value. Determining a physical process responsible for the
non-zero eccentricity is outside the scope of this work.\\
\indent The inclination values reported in this work are measured with respect
to the J2000 equatorial frame (i.e., with respect to the Earth's
equatorial plane). The fits provide no clues to the orientation
of Dysnomia's orbit with respect to Eris' equatorial plane, they only
provide a measurement of the orientation of the orbit pole with
respect to Eris' heliocentric orbit. If the
orbit pole and Eris' rotation pole were aligned, this would be
evidence for a giant impact formation scenario for Dysnomia, and would
enable the determination of seasons on Eris (as well as Dysnomia,
assuming its rotation pole is parallel to the orbit and Eris
rotation poles). However, there is no indication that this is the
case, and estimates of the tidal damping timescale for the inclination
are exceedingly long. The ratio of inclination and eccentricity
damping timescales is given by Murray and Dermott (1999):
\begin{equation}
\frac{\tau_i}{\tau_e}=7\left(\frac{\textrm{sin}i_0}{\textrm{sin}\epsilon}\right)^2\left(\frac{1}{\textrm{cos}i_0}\right),\nonumber
\end{equation}
where $i_0$ is the initial inclination of the system and $\epsilon$ is
the angle between the tidal bulge and the line connecting the centers
of the primary and secondary. It is related to the tidal dissipation
factor, $Q$, by $\epsilon=\frac{1}{2Q}$. Thus, for
$Q$=100, $\epsilon$=0.005 radians. For an initial inclination of
1$^{\circ}$, the ratio of timescales is $\sim$85, so for the
optimistic case of an eccentricity damping timescale equal to 5 Myr,
the inclination damping timescale is 425 Myr. This timescale increases
quickly with initial inclination, so that for a modest initial
inclination of only 3.27$^{\circ}$
the inclination damping timescale is comparable to the age of the
solar system. If Dysnomia is a captured satellite, the initial
inclination of its orbit could have been anywhere between 0$^{\circ}$
and 90$^{\circ}$ in order to match the constraint from this work that
Dysnomia's orbit is prograde. The range of initial inclinations that
result in aligned orbit and spin poles is small and there is no
preference for an initial low-inclination orbit in this scenario,
which suggests that if Dysnomia is a captured satellite, it would currently be on
an inclined orbit with respect to Eris' equatorial plane.\\
\indent The inclination damping timescale considered above only takes
into account the effects of tidal bulges but not the oblateness of the
primary.  Even a modest oblateness of 1\%, the upper limit for Eris
determined by Sicardy et al. (2011), is enough to significantly alter
a satellite's orbital evolution (e.g., Porter and Grundy,
2012). Therefore, additional modeling work is needed to evaluate the
evolution of Dysnomia's orbit, particularly the inclination with
respect to Eris' equatorial plane. This investigation, which is outside the
scope of the current work, would need to account for the
effects of an oblate Eris on the damping of the orbit's inclination,
since tidal damping alone is clearly negligible.\\
\indent In Table 2, we note the high $\chi^2$ for Epoch 2 and the
combined orbit fits compared to the Epoch 1 orbit fit. This is easily
explained by the difference in the size of the error bars between the
two epochs (Figure 2), and the smaller error bars for Epoch 2
weight the combined fit towards those points. However, the
high $\chi^2$ values also indicate that a Keplerian orbit is not the
best-fit to the Epoch 2 measurements. In fact, a Keplerian orbit can be excluded
at the 5.8-$\sigma$ and 6.3-$\sigma$ levels for the Epoch 2 and combined fits,
respectively. Possible physical explanations for a non-Keplerian orbit include
precession of Dysnomia's orbit due to an oblate Eris; a non-spherical
Dysnomia; the presence of an additional, previously undetected
satellite; or center-of-light versus center-of-body (CoL-CoB) offsets due
to large-scale albedo patterns. Evaluation of the effects of a potential
tidal bulge on Eris is the subject of future work and is outside the
scope of this particular paper. A discussion of additional
satellites around Eris was initially presented in Murray et al. (2018), with
more detailed work currently in preparation.}

\section{Summary}
\doublespacing{We used relative astrometry from WFC3/HST images obtained
  in January and February 2018, combined with previously published and
  unpublished HST and Keck data, to compute a new orbit for Dysnomia and
  break the degeneracy in the orbit pole orientation. Highlights of the
  results and interpretations include:
\begin{itemize}
\item The calculation of a new orbital period for Dysnomia,
  15.785899$\pm$0.000050 days, which agrees with the value from Brown
  and Butler (2018) to within $\sim$3 seconds. Both investigations
  made use of the relative astrometry from Brown and Schaller (2007),
  yet the more recent periods are not in agreement with those reported
  in Brown and Schaller (2007) for Orbit 1 (15.772$\pm$0.002 days) or
  Orbit 2 (15.774$\pm$0.002 days). In other words, the older results
  are now outliers and have never been replicated, even when
  considering the same data set.
\item An orbit pole obliquity of 78.29$\pm$0.65$^{\circ}$, which
  agrees with Orbit 1 of Brown and Schaller (2007). The orbit opening
  angle in 2018 was 42$^{\circ}$ and the next mutual events
  season will be in 2239. Dysnomia's orbit is prograde.
\item An update to the system density, 2.43$\pm$0.05 g
  cm$^{-3}$, which takes into account the system mass from the
  combined fit of (1.6466$\pm$0.0085)$\times$10$^{22}$ kg and volumes
  of both objects. If Eris and Dysnomia have the same
  density, Dysnomia accounts for $\sim$3\% of the system mass.
\item A non-zero eccentricity for Dysnomia's orbit,
  0.0062$\pm$0.0010, which is reported at a significance of
  6.2-$\sigma$. Tidal circularization should occur within $\sim$17 Myr
  even if Dysnomia's density is a third of the system
  density. Explanations for the non-zero eccentricity involving center-of-light
  versus center-of-body offsets or systematic errors are not favored;
  determining a physical cause of this non-zero eccentricity is
  outside the scope of this work.
\item A Keplerian orbit for the combined fit that can be excluded at the
  6.3-$\sigma$ level, suggesting precession of Dysnomia's orbit due to
  the oblateness of Eris, an irregularly shaped Dysnomia, an unseen
  interior satellite, or center-of-light versus center-of-body offsets.
\end{itemize}
A possible future investigation related to this work is an evaluation of
the non-Keplerian nature of Dysnomia's orbit in the context of an oblate
Eris. The same investigation could provide clues to the
inclination of Dysnomia's orbit with respect to Eris' equatorial
plane, which could in turn enable an evaluation of short- and
long-term seasonal cycles on Eris. A future paper will
provide a detailed discussion on the search for a Pluto-like minor
satellite system around Eris.}

\section*{Acknowledgements}
\doublespacing{The authors would like to thank the two anonymous
  reviewers for their helpful suggestions as well as
  Darin Ragozzine and Leslie Young for their constructive
  discussions. The authors appreciate the work of Crystal Mannfolk, Linda
  Dressel, and Kailash Sahu of STScI in helping to optimize the HST
  observations prior to execution. This work is based on observations made with
  the NASA/ESA Hubble Space Telescope, obtained from the data archive
  at the Space Telescope Science Institute. STScI is operated by the
  Association of Universities for Research in Astronomy, Inc., under
  NASA contract NAS 5-26555. Support for this work was provided by
  NASA through grant number GO-15171.001 from STScI.}

\section*{References}
\begin{hangparas}{0.25in}{1}
\doublespacing{
\noindent Bannister, M.T., et al., submitted. Expanding Horizons: The
need for direct exploration of the diverse trans-Neptunian Solar System.

Barucci, M.A., et al., 2011. New insights on ices in Centaur and
transneptunian populations. Icarus 214, 297-307.

Brown, M.E., et al., 2006. Direct measurement of the size of 2003
UB313 from the Hubble Space Telescope. ApJ 643, L61-L63.

Brown, M.E., Schaller, E.L., 2007. The mass of dwarf planet
Eris. Science 316, 1585.

Brown, M.E., 2012. The compositions of Kuiper Belt
objects. Ann. Rev. Earth and Plan. Sci., 40, 467-494.

Brown, M.E., Butler, B.J., 2018. Medium-sized satellites of large
Kuiper Belt objects. AJ 156, 164.

Carraro, G., et al., 2006. Time series photometry of the dwarf planet
Eris (2003 UB313). A\&A 460, L39-L42.

Duffard, R., et al., 2008. A study of photometric variations on the
dwarf planet (136199) Eris. A\&A 479, 877-881.

Fraser, W.C., et al., 2017. All planetesimals born near the Kuiper
Belt formed as binaries. Nature Astronomy 1, 0088.

Gladman, B., Marsden, B.G., Vanlaerhoven, C., 2008. Nomenclature in
the outer solar system. In: Barucci, M.A., Boehnhardt, H., Cruikshank,
D.P., Morbidelli, A. (Eds.), The Solar System Beyond
Neptune. University of Arizona Press, Tucson, 43-57.

Goldreich, P., Soter, S., 1966. Q in the solar system. Icarus 5, 375-389.

Gomes, R.S., 2003. The origin of the Kuiper Belt high-inclination
population. Icarus 161, 404-418.

Hainaut, O.R., Boehnhardt, H., Protopapa, S., 2012. Colours of minor
bodies in the outer solar system. II. A statistical analysis
revisited. A\&A 546, A115.

Krist, J., 1993. Tiny Tim: An HST PSF simulator. In: Hanisch, R.J.,
Brissenden, R.J.V., Barnes, J. (Eds.), ASP Conf. Ser. 52, Astronomical
Data Analysis Software and Systems II. ASP, San Francisco, pp. 536.

Lacerda, P., et al., 2014. The albedo-color diversity of
transneptunian objects. ApJL 793, L2.

Levison, H.F., Morbidelli, A., 2003. The formation of the Kuiper Belt
by the outward transport of bodies during Neptune's migration. Nature
426, 419-421.

Levison, H.F., et al., 2008. Origin of the structure of the Kuiper
Belt during a dynamical instability in the orbits of Uranus and
Neptune. Icarus 196, 258-273.

Malhotra, R., 1993. The origin of Pluto's peculiar orbit. Nature 365,
819-821.

Malhotra, R., 1995. The origin of Pluto's orbit: Implications for the
solar system beyond Neptune. AJ 110, 420-429.

M{\"u}ller, T.G., et al., 2010. ``TNOs are Cool'': A survey of the
trans-Neptunian region. I. Results from the Herschel science
demonstration phase (SDP). A\&A 518, L146.

Murray, C.D., Dermott, S.F., 1999. Solar System Dynamics. Cambridge
University Press.

Murray, K., Holler, B.J., Grundy, W., 2018. Search for a Pluto-like
satellite system around Eris. AAS/DPS 50, \#311.08.

Noll, K.S., et al., 2008. Evidence for two populations of classical
transneptunian objects: The strong inclination dependence of classical
binaries. Icarus 194, 758-768.

Noll, K.S., Parker, A.H., Grundy, W.M., 2014. All bright cold
classical KBOs are binary. AAS/DPS 46, \#507.05.

Parker, A.H., et al., 2016. Discovery of a Makemakean moon. ApJL 825,
L9.

Porter, S.B., Grundy, W.M., 2012. KCTF evolution of trans-neptunian
binaries: Connecting formation to observation. Icarus 220, 947-957.

Roe, H.G., Pike, R. E., Brown, M.E., 2008. Tentative detection of the
rotation of Eris. Icarus 198, 459-464.

Sicardy, B., et al., 2011. A Pluto-like radius and a high albedo for
the dwarf planet Eris from an occultation. Nature 478, 493-496.

Spencer, J.R., Denk, T., 2010. Formation of Iapetus' extreme albedo
dichotomy by exogenically triggered thermal ice migration. Science
327, 432.

Stern, S.A., et al., 2015. The Pluto system: Initial results from its
exploration by New Horizons. Science 350, aad1815.

}
\end{hangparas}

\end{document}